\documentstyle[prd,aps]{revtex}
\input epsf.tex
\def\DESepsf(#1 width #2){\epsfxsize=#2 \epsfbox{#1}}
\begin{document}
\draft
\twocolumn[\hsize\textwidth\columnwidth\hsize\csname
@twocolumnfalse\endcsname
\preprint{CTP-TAMU--98}
\title{The Tevatron Tripler: \\How to Upgrade the Fermilab Tevatron for the Higgs
Boson and Supersymmetry}
\author{ P. McIntyre, E. Accomando, R. Arnowitt, B. Dutta, T. Kamon, and  A.
Sattarov}
\address{Department of Physics, Texas A \& M University, College Station, TX 
77843}
\date{August, 1999}
\maketitle
\begin{abstract}

Recent advances in superconductor properties and superconducting magnet 
technology have made it possible to build cost-effective, high-performance 
dipoles with a field of 12 Tesla - 3 times the field strength of the Tevatron.  
Such magnets could be used to upgrade Fermilab's collider in its existing tunnel 
to a collision energy $\sqrt s$= 6 TeV and luminosity {$\cal L$}$>$ 10$^{33}$
cm$^{-2}$s$^{-1}$.   We have calculated the parton luminosities for
quark-antiquark and gluon-gluon  scattering for the Tevatron, the Tripler, and
LHC.   In most models of the Higgs field and supersymmetry,  the Tripler would
have a high likelihood to discover  many of the predicted particle states.
\end{abstract}   

\pacs{ }

 \vskip2pc]
Accelerator technology has paced the discovery of new particles at hadron
colliders.   With the invention of beam cooling techniques\cite{1,2},  it became
possible to make colliding beams of protons and antiprotons in existing
accelerators \cite{3}. ${\bar p}p$ colliding beams make use of the valence antiquarks
within the antiproton to access the annihilation channels  through which many new
particles are produced most directly.  This development led to the discovery of
the weak bosons \cite{4}.
 With the development of superconducting accelerator magnets \cite{5},
  it became possible to extend the energy reach in collisions to 2 TeV, 
 leading to the discovery of the top quark \cite{6}. 
 Continuing progress towards higher collision energy for hadron collisions 
 requires the development of magnets with yet higher magnetic field strength.  Three
properties of high-field dipoles have so far inhibited such progress: the
intrinsic performance of the superconductor, degradation of the superconductor
by mechanical stress, and protection of the magnet in the event of quench.

\section{Optimization of high-field dipoles}
 The alloy NbTi has been used as the
superconductor for all superconducting colliders: Tevatron, HERA, RHIC, and now
LHC \cite{7}.
 This alloy makes a very robust cable and has been perfected to a
high-performance technology.  
 Unfortunately its upper critical field limits the maximum field that can be
produced, so that even cooling with 
 superfluid helium (1.8$^0$K) the 8.4 Tesla field of the LHC is a practical
limit (Figure 1). 

Nb$_3$Sn has recently matured as a practical superconductor,
so that strand is now available with a current density  j$_{\rm sc}$ = 2,000
A/mm$^2$ at 12 Tesla \cite{8}.  Bi-2212 strand is now reaching 1,000 A/mm$^2$, 
with an upper critical field above any practical field strength.   While these
new conductors in principle offer much higher field strength for dipoles,  they
are brittle materials and undergo strain degradation of supercurrent capacity
j$_c$ at  moderate stress levels.  Figure 2 shows the dependence of j$_c$ on stress
in the superconducting  filaments of multi-filament Nb$_3$Sn strands made by three
processes - powder-in-tube (PIT), bronze,  and modified jelly roll (MJR).  Only
the MJR process yields high j$_c$(0), but it exhibits the largest  degradation from
strain, so that coil stress must be limited to $\sim$120 MPa.  Bi-2212 is even more
brittle,  degrading at stresses of $\sim$70 MPa.  The stress from the Lorentz forces
in a  dipole coil accumulates from the inside to the outside of the coil.   For
a central field of 12 Tesla, the peak stress in a conventional multi-shell
cos$\theta$ coil would reach $\sim$150 MPa. 

A technique of stress management was devised
recently, in which a support matrix is integrated within the coil\cite{9}. 
 The matrix of high-strength ribs and plates intercepts the stresses acting on
the inner coil elements and bypasses them past 
 the outer elements, so that stress cannot accumulate beyond $\sim$70 MPa.  Since the
matrix material (e.g. Inconel) can support 
 much greater loading stress than the fragile coils, this approach effectively
removes mechanical stress as a limit for 
 high-field magnet technology. 
 
 A remaining limit to high-field dipole technology
is the requirement that the coil not be damaged in the event that  it loses its
superconducting state (quench). The coil consists of an array of windings, each
a rectangular (Rutherford)  cable containing $\sim$30-40 superconducting
strands. Each strand in turn contains $\sim$300-1000  filaments of
superconductor drawn within a copper matrix.  The main cost in making
superconducting strand  is the complex sequence of stacking, extruding, drawing,
and heat treatment needed to achieve optimal superconducting performance.  
Since the cost scales with the volume of processed strand, {\it the added copper
costs as much as the superconductor itself}.

Figure 1 shows the critical current vs. field in the filaments of several
practical superconductors.   The magnetic field in the superconducting coil is
largest in its inner regions, and decreases and even reverses direction in its
outer regions.   In an optimized series coil, the inner windings would have
smaller cross-section, and the outer windings would have larger cross-section.  
This optimization is called grading, and optimally uses the available current
density j$_c$(B) throughout the coil.

When a coil quenches, however, the current still must be carried in the quenched
regions of each winding.   Since superconducting alloys are quite resistive in
the normal state, a substantial amount of a good conductor  (e.g. copper) must
be integrated into the cables. Copper is required for two reasons.   A modest
amount of copper is needed to {\it stabilize} the superconducting filaments so
that when a single  filament quenches, the surrounding copper can carry the
current and conduct heat from the quenched filament  sufficiently that the
filament can regain its superconducting state.  Quench stability for high field
Nb$_3$Sn coils typically
requires  an amount of copper equal to about 40$\%$ of the strand cross-section.

Copper is also required to {\it protect} the coil in the event that an entire
region of cable begins to quench.   During such quench the current must be
transferred to the copper so that heat dissipation  can be limited sufficiently
that the stored magnetic field energy can be safely dissipated  throughout the
cold mass of the magnet.  This condition requires that sufficient copper be 
distributed in the cable so that the current density in copper during quench is
j$_{\rm Cu}<$ 1,500 A/mm$^2$. 

Conventionally all of the copper for both stability and quench protection is
extruded as a matrix within  each multi-filament strand.  We recently devised an
alternative strategy \cite{10}, in which only the copper  for quench stability
is integrated within the strand, but the (typically much larger amount of) 
copper for quench protection is provided by cabling pure copper strands together
with copper-stabilized  superconducting strands in each cable.  The proportion
of copper and superconducting strands  is chosen to provide the necessary quench
protection in each region of the coil.

Figure 3 shows the design of a 12 Tesla dipole that contains provisions for 
stress management and also an optimized arrangement of copper and superconductor 
in the cables.  The inner segment of the coil uses cable containing two
superconducting strands
 for each copper strand.  The middle segment of the coil uses cable containing
equal numbers 
 of copper and superconducting strands.  The magnetic field in the outer segment
is everywhere less than 
 7.2 Tesla, and so high-performance NbTi cable can be used there
 (half of the entire coil).

The impact of this design approach can be appreciated by comparing the coil 
area required for conventional dipoles and for designs using the above
innovations.   Figure 4 shows the coil cross-section area for each of the NbTi
cos$\theta$ dipoles of  existing colliders, and for three designs using the
above design strategy: the 12 Tesla Tripler dipole of Figure 3 (only the Nb$_3$Sn
 coil area is counted for this purpose), a 15 Tesla all-Nb$_3$Sn dipole, and
finally a 20 Tesla dipole using Bi-2212 inner coils and 
 Nb$_3$Sn outer coils.  These last two designs have reduced aperture (4 cm) and
may be appropriate for a future ultimate-energy hadron collider (VLHC).   The
amount of superconductor required for conventional designs increases
quadratically  with field, while the superconductor required for the optimized
designs increases linearly.   The high-field designs require far less
superconductor/TeV than do dipoles for the SSC and LHC.

\section{Tevatron Tripler}
 The new dipole design strategy makes it possible to upgrade
Fermilab's Tevatron by replacing its ring of 4 Tesla dipoles by a ring of 12 Tesla
dipoles.   In addition to the high-field dipoles, the upgraded ring requires
quadrupoles with 3 times the gradient of those used in the Tevatron.   A
block-coil Nb$_3$Sn quadrupole with the necessary gradient has been designed.
  
  The new magnets could be installed on the existing stands
  that until recently supported the original Main Ring.  The Tripler would
make colliding beams with a collision energy $\sqrt s$ = 6 TeV.  The
antiproton source, injectors, cryogenics, RF systems and detectors of the
Tevatron could all be used with modest upgrade. 

The luminosity of the Tripler is 
\begin{eqnarray}L&=&{{N_bN_pN_{\bar
p}f}\over{\beta^*\sqrt{\epsilon_h\epsilon_\nu}}}\gamma=10^{33}{\rm cm}^{-2}{\rm
s}^{-1}
\end{eqnarray} where N$_b$ = 160 is the number of bunches,
N$_p$=3$\times$10$^{11}$ and N$_{\bar p}$= 4$\times$10$^{10}$  are the numbers of
particles per bunch, f = 50 kHz is the revolution frequency, $\beta^*$= 0.25 m 
is the focal length at the collision point, and $\epsilon_h$ and $\epsilon_\nu$
are the invariant emittances  in the horizontal and vertical phase space of the
beam.  The luminosity scales with the relativistic $\gamma$ of the beam. 
 All of the other parameters are the same for the Tripler as they will be for the
Tevatron as that being upgraded for its next run.
  
The Tripler must accommodate counter-circulating beams of protons and
antiprotons.   For high-luminosity collisions, the beams must be fully separated
everywhere except at  the two locations where they collide inside the detectors,
so that beam-beam tune shift  is minimized.  This is accomplished using
electrostatic deflectors, so that the two beams  trace a double helix within the
beam tube of the dipoles.  The 6 cm aperture of the design  in Figure 3 is
sufficient for this purpose.  

A last requirement is that the field distribution
in the dipoles be sufficiently uniform so  that high luminosity can be
maintained during a days-long store.  This requirement is usually  expressed by
requiring that the multipoles b$_n$ should be smaller than $\sim10^{-4}$
cm$^{-n}$.  We achieve  this criterion by current-programming the inner
windings.  Figure 5 shows the multipoles vs.  field over the 20:1 range from
injection to collision energy. At 12 Tesla, synchrotron radiation is not yet
enough to require the use of an  intermediate-temperature liner in the beam
tube.  The total radiated power in the ring is 
\begin{eqnarray}P&=&{2e^6c^2\over{9\epsilon_0(m_pc^2)^4}}E^2B^2N_b(N_p+N_{\bar
p})=400 W
\end{eqnarray} This power could be removed using the installed cryogenic
refrigeration capacity in the Tevatron.

\section{Parton luminosities} In order to assess the potential of the Tripler
for the discovery of new particles,  we have calculated parton luminosities with
which the constituents of colliding hadrons interact.   Parton luminosities are
calculated for $u\bar d$ scattering and for gluon-gluon scattering, for the 
cases of the Tevatron, the Tripler, and the LHC.   The calculations use the
CTEQ4 parton distributions \cite{11}.   The results are presented in Figure 6 as a
function of $\sqrt{\hat s}$, the c.m. energy for the colliding partons. 
 The signals for the Higgs boson and for supersymmetry (SUSY) at a hadron
collider sort into two 
 broad categories: those with signatures of multiple leptons and missing energy,
for which the 
 cross-sections are small but the signal/background ratio is fairly large; and
those with signatures 
 of multiple quarks (jets) and missing energy, for which the cross sections are
larger but the signal/background ratio is small.  
 The multi-lepton states are accessed primarily through quark-antiquark
annihilation, 
 while the multi-quark states are accessed primarily through gluon fusion.  
 Searches for Higgs and SUSY states at the Tevatron concentrate primarily on the
multi-lepton signatures, 
 taking advantage of the valence antiquark content of the antiproton that
enhances the high-mass spectrum of 
 quark-antiquark annihilation (Figure 6).  The preparation for similar searches
at LHC concentrates on 
 the multi-quark signatures, since for pp collisions an antiquark must be drawn
from the sea with reduced $\hat s$.

To evaluate the signals and backgrounds for the several signatures for  Higgs
and SUSY as a function of mass scale, one must conduct a detailed  Monte Carlo
generation of events and simulation of detector cuts and  acceptances.  This
work is in progress.   A first estimate of sensitivity can be obtained by
extrapolating  the fully simulated signals and backgrounds for each signature
that  have been done by the CDF and D0 collaborations in preparation for  Run II
of the Tevatron \cite{12}. The yields of signal events are scaled  from the
Tevatron simulation, using the integrated luminosities for scattering  from
Figure 6 for parton c.m. energies greater than $M_H+M_Z$.   Similarly the
background events are scaled from the Tevatron simulation  using the gluon-gluon
luminosities.  Table 1 presents the signals  and background for a 20 fb$^{-1}$
data sample (a one-year run at the Tripler  luminosity).  Also presented are the
$\chi^2$ that would result from a  Baysean combination of the several
signatures. The 95$\%$ confidence limit  would extend to 400 GeV/c$^2$ mass
scales.  

In previous studies, SUSY \cite{13} and Higgs \cite{14} signals were  evaluated
for a proposed 4 TeV upgrade of the Tevatron.   Extrapolating from those
results, the Tripler should be sensitive to  gluinos and squarks with masses up
to $\sim$750 GeV, and would be able  to see the trilepton signal (associated
production of chargino and  neutralino) over much of the interesting parameter
space for both universal  and non-universal soft breaking.
\section{Conclusion} The new developments in superconducting dipole design make
it feasible to  extend the high-field frontier for future hadron colliders.  
The Tripler is a first opportunity to use this technology to advantage, 
extending the reach for Higgs and SUSY particles to $>$500 GeV/c$^2$.  Its
extension of the reach for quark annihilation processes is complementary to the
primarily gluon-mediated processes at LHC, and the lepton annihilation channels
at NLC.  For this reason the Tripler would remain a competitive facility
through much of the physics life of LHC and NLC. 

It is a pleasure to acknowledge
stimulating discussions with W. Barletta and R. Scanlan of LBNL, and R. Noble of
Fermilab.  This work was supported by DOE grant DE-FG03-95ER40924, NSF grant
PHY-9722090, and by a Texas Advanced Technology Program grant.

Table 1. Signal and background for Higgs signatures at
Tripler.
\begin{center}
 \begin{tabular}{|c|c|c|c|c|}  \hline
 {\rm Signature}&\multicolumn{4}{c|}{\rm Signal (Background) in 20 $fb^{-1}$}
 \\&\multicolumn{4}{c|}{\rm Higgs mass (GeV/c$^2$)}\\
&\multicolumn{1}{c}{150}&\multicolumn{1}{c}{200}&\multicolumn{1}{c}{300}&
\multicolumn{1}{c|}{500}\\\hline
 $l^{\pm}l^{\pm}l^{\pm}$&5(26)&2(6)&1(2)&0.3(0.3)\\
 $l^{\pm}l^{\pm}\nu\nu$&274(31K)&116(8K)&55(2.4K)&18(400)\\
  \hline
  $l^{\pm}l^{\pm}jj$&86(1.3K)&36(315)&17(100)&5.6(160)\\\hline
  $\chi^2$&9&7&5&3\\\hline
\end{tabular}
\end{center}
\begin{figure}
\vspace{1 cm}
\centerline{ \DESepsf(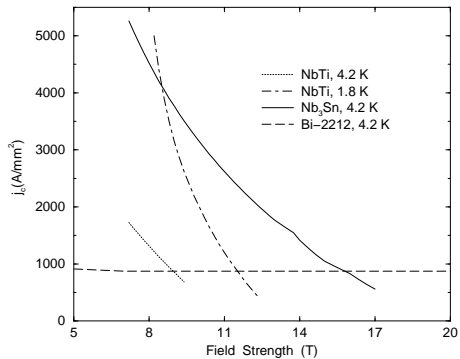 width 6 cm) }
\smallskip
\caption {Field dependence of critical current density.}
\end{figure}
\begin{figure}
\vspace{1 cm}
\centerline{ \DESepsf(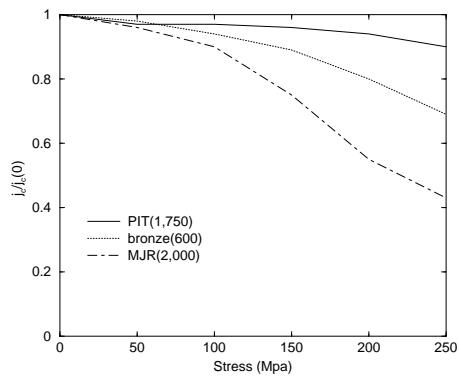 width 6 cm) }
\smallskip
\caption {Strain degradation of Nb$_3$Sn superconductors.}
\end{figure}
\begin{figure}
\vspace{1 cm}
\centerline{ \DESepsf(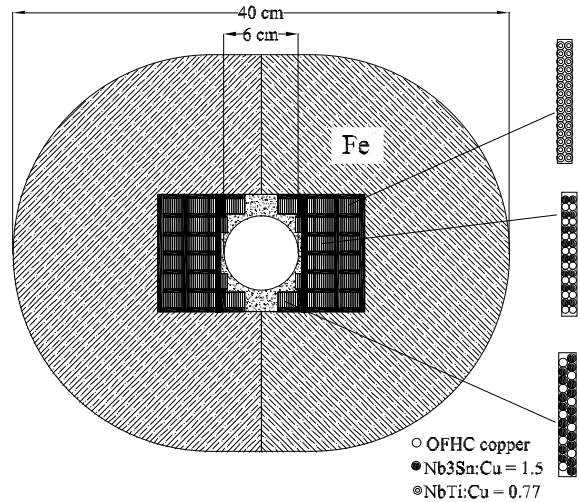 width 8 cm) }
\smallskip
\caption {Tesla dipole suitable for use in a Tevatron Tripler. Cross sections
of the Rutherford coil in each section are shown.}
\end{figure}
\begin{figure}
\vspace{1 cm}
\centerline{ \DESepsf(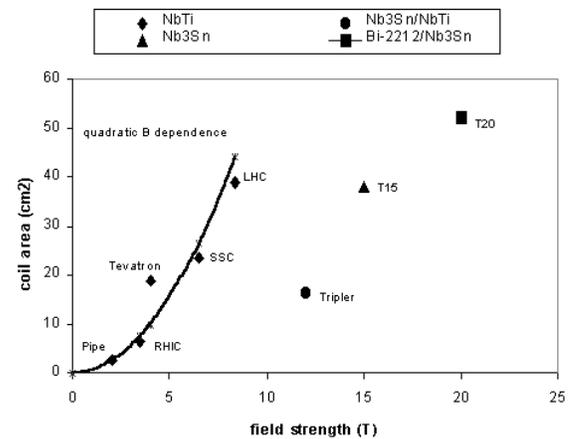 width 8 cm) }
\smallskip
\caption {Coil area for conventional dipoles and for designs using
 stress management and superconductor optimization.}
\end{figure}\newpage
\begin{figure}
\vspace{1 cm}
\centerline{ \DESepsf(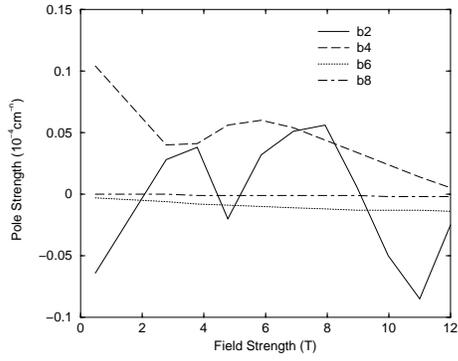 width 6 cm) }
\smallskip
\caption {Multipole moments after current programming.}
\end{figure}
\begin{figure}
\vspace{1 cm}
\centerline{ \DESepsf(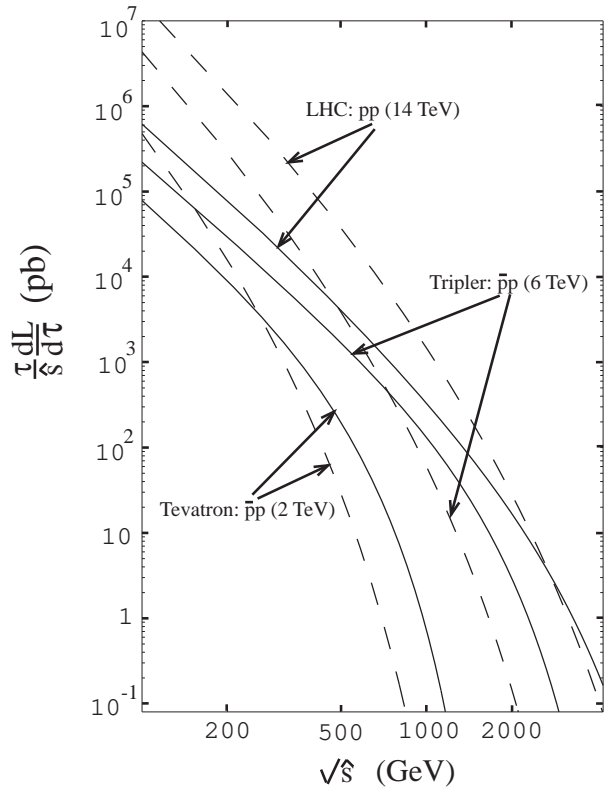 width 8 cm) }
\smallskip
\caption {Parton luminosities for $u{\bar d}$ (solid lines) and gg scattering
(dashed lines).}
\end{figure}
\end{document}